\documentclass[twocolumn,floatfix, nofootinbib,prd,preprintnumbers,showpacs,showkeys,superscriptaddress,longbibliography]{revtex4-2}

\usepackage{graphicx} 
\usepackage{orcidlink}
\usepackage[top=2cm,bottom=2cm,left=1cm,right=1cm,marginparwidth=1.75cm]{geometry}
\title{Asy Static Finite Distance}
\usepackage{amssymb}
\usepackage{amsmath}
\date{\today}
\usepackage{mathtools}
\usepackage{comment}

\usepackage[mathscr]{euscript}
\usepackage{amsmath}
\usepackage{graphicx}
\usepackage{dcolumn}
\usepackage{bm}
\usepackage{epsfig}
\usepackage{amssymb,latexsym,mathrsfs}
\usepackage{graphicx}
\usepackage{color}
\usepackage{hyperref}
\usepackage{float}
\usepackage{diagbox}

\usepackage{textgreek}

\usepackage{tikz}

\usepackage{amsmath}
\usepackage{physics}
\usepackage{csquotes} 
\newcommand*\diff{\mathop{}\!\mathrm{d}}

\hypersetup{
    colorlinks=true,
    linkcolor=red,
    citecolor=blue,
}

\usepackage{amsmath}
\usepackage{amssymb}
\usepackage{subfigure}
\usepackage{hyperref}
\usepackage{url}
\usepackage{xcolor}
\usepackage{color}
\definecolor{amaranth}{rgb}{0.9, 0.17, 0.31}
\definecolor{purple(munsell)}{rgb}{0.62, 0.0, 0.77}
\definecolor{americanrose}{rgb}{1.0, 0.01, 0.24}
\definecolor{palatinateblue}{rgb}{0.15, 0.23, 0.89}
\definecolor{royalblue(web)}{rgb}{0.25, 0.41, 0.88}
\definecolor{hanpurple}{rgb}{0.32, 0.09, 0.98}
\definecolor{beaublue}{rgb}{0.74, 0.83, 0.9}
\definecolor{carminered}{rgb}{1.0, 0.0, 0.22}
\definecolor{brightpink}{rgb}{1.0, 0.0, 0.5}
\definecolor{vividviolet}{rgb}{0.62, 0.0, 1.0}
\hypersetup{ linktoc=all,
    colorlinks, linkcolor={palatinateblue},
    citecolor={brightpink}, urlcolor={amaranth}}

\usepackage{comment}

\newcommand{\be}{\begin{equation}}
\newcommand{\ee}{\end{equation}}
\newcommand{\bs}{\begin{split}} 
\newcommand{\bea}{\begin{eqnarray}}
\newcommand{\eea}{\end{eqnarray}}

\newcommand{\mike}[1]{\textcolor{red}{[{\bf MG}: #1]}}

\newcommand{\bes}{\begin{subequations}}
\newcommand{\ees}{\end{subequations}}

\newcommand{\bo}{\raise-1mm\hbox{\Large$\Box$}}

\begin{document}

\title{Classical Acceleration Temperature (CAT) in a Box}
\author{Ahsan Mujtaba\,\orcidlink{0009-0006-2595-6991}}
\affiliation{Physics Department, NED University of Engineering \& Technology,\\ Karachi, 75270, Pakistan}
\affiliation{Physics Department, Lahore University of Management Sciences (LUMS),\\ Lahore 54792, Pakistan\\}
\author{Maksat Temirkhan\,\orcidlink{0000-0001-6283-3401}}
\affiliation{Physics Department \& Energetic Cosmos Laboratory, Nazarbayev University,\\
Astana 010000, Qazaqstan}
\author{Yen Chin Ong\,\orcidlink{0000-0002-3944-1693}}
\email{ycong@yzu.edu.cn}
\affiliation{Center for Gravitation and Cosmology, Yangzhou University,\\
Yangzhou 225002, China}
\affiliation{Shanghai Frontier Science Center for Gravitational Wave Detection, Shanghai Jiao Tong University,\\ Shanghai 200240, China}
\author{Michael R.R. Good\,\orcidlink{0000-0002-0460-1941}}
\email{michael.good@nu.edu.kz}
\affiliation{Physics Department \& Energetic Cosmos Laboratory, Nazarbayev University,\\
Astana 010000, Qazaqstan}
\affiliation{Leung Center for Cosmology and Particle Astrophysics,
National Taiwan University,\\ Taipei 10617, Taiwan}

\begin{abstract} 

A confined, slow-moving, accelerating electron is shown to emit thermal radiation.  Since laboratories face spatial constraints when dealing with rectilinear motion, focusing on a finite total travel distance combines the benefits of simple theoretical analysis with prospects for table-top experimentation. We demonstrate an accelerated moving charge along an asymptotically static worldline with fixed transit distance and slow maximum speed, emitting self-consistent analytic power, spectra, and energy. The classical radiation is Planck distributed with an associated acceleration temperature. This is the first fully parametrized, spectrum-solved, finite-distance worldline.
\end{abstract}
\maketitle
\tableofcontents
\section{Introduction}

Quantum radiation from thermal black holes \cite{Hawking:1974sw} is studied in analog by quantum radiation from moving mirrors \cite{DeWitt:1975ys,Davies:1976hi,Davies:1977yv}. Ongoing studies exploit the similarities between these systems in investigating quantum radiation \cite{Ievlev:2023ejs,Osawa:2024fqb,Lin:2021bpe,Kumar:2023kse,Reyes:2021npy}. Interestingly, classical radiation from a moving point charge has a one-to-one map to the quantum radiation from a moving mirror, see e.g., \cite{Nikishov:1995qs,Ritus:2022bph,Ievlev:2023akh}.  Recently, the accelerating electron from beta decay has been experimentally shown to emit classical thermal radiation commensurate with infinite particle production \cite{Ievlev:2023inj, Good:2022eub,ptep}.    

A completely evaporated black hole should release a finite amount of particles (IR-finite) \cite{Ievlev:2023xzv}.  A moving mirror must return to rest for finite particle emission.  It is a strong constraint on the motion to return to rest.  So it is no surprise that rectilinear motions possessing asymptotic rest with solved Bogolyubov coefficients have provided insight into the general classical-quantum character of particle creation and acceleration radiation \cite{Walker_1982,Good:2017kjr,Good:2018aer,Good:2023ncu,Good:2019tnf,Good:2020fsw,Good:2023ncu,Moreno-Ruiz:2021qrf,Ievlev:2023xzv,Ievlev:2023akh}.  Perhaps the primary benefits to exploring these trajectories have been the finite energy emission and finite particle production. Nevertheless, asymptotically resting worldlines also suffer no information loss (unitary) and are free from IR divergences. 

Asymptotic rest pays the cost of quantum purity \cite{wilczek1993quantum, Good:2019tnf} and is a physically well-motivated way to preserve the unitary interaction between the mirror and its quantum field \cite{PisinChen2017}. In the context of the semi-classical flying mirror (the mirror is classical with a position and velocity simultaneously known while the fields are quantized), information loss occurs when a global acceleration horizon is formed. With asymptotic rest, late-time light rays will never evade the mirror and cannot descend past a horizon into its analog black hole. 

Of course, quite independently of its utility as a tool to model Hawking evaporation of black holes, moving mirror solutions are interesting in and of themselves. Finding analytic solutions exhibiting certain properties one wishes to study is generally difficult. If we want to move beyond theoretical toy models to laboratory experiments, we would like to construct a solution whose trajectory is bounded in space since a laboratory is finite in size. 
The problem is that, in the literature, only one asymptotically static mirror is known to travel a finite distance; see Table \ref{dualitytable}.   This trajectory (`Arctx') cannot be simultaneously solved for its Bogolyubov coefficients and parametrized in terms of maximum speed \cite{good2013time}.  Thus, studying how its maximum speed relates to its particle spectrum is challenging. Likewise, it is unknown whether or not finite travel affects particle creation.

Faced with these intractabilities, we approach these issues by investigating a new solution that can be simultaneously solved and parametrized.  We introduce the first spectrum-solved finite-distance worldline, expressed in terms of its maximum speed.  The trajectory facilitates an examination of confinement on particle spectrum and count.  The result is Planck-distributed thermal radiation.  

The structure of the paper is as follows.
In Sec.~\ref{sec:electronbox}, we review the essential dynamics for the trajectory, developing an intuition for the power and the energy emission. Here, we reveal the interesting physics of `big-box' energy production. Sec.~\ref{sec:spectrum} is devoted to spectral analysis of the radiation, running into intractable generality. Nevertheless, we find a spectrum for a specific box size and demonstrate consistency with the energy emission.  With the spectrum in hand, we compute effective temperature functions based on the trajectory dynamics in Sec.~\ref{sec:CAT} and demonstrate thermal radiation.  In Sec.~\ref{sec:particles}, we use the spectral results to draw several conclusions about the photon particle production. Sec.~\ref{sec:concl} summarizes the main findings, emphasizing the radiation is thermal from a finite travel region, i.e., a CAT in the box. Units are $c = \mu_0 = \epsilon_0 =1$. 

\begingroup
    \renewcommand{\arraystretch}{2}
\begin{table}[ht] 
\centering
\begin{tabular}{|>{\centering\arraybackslash}m{4.8cm}|>{\centering\arraybackslash}m{4.2cm}|}
\hline
Asymptotic Static Motions & Distance Travelled \\
\hline\hline
Walker-Davies \cite{Walker_1982} & $\infty$ \\\hline
{\bf Arctx} \cite{good2013time} & finite \& unparametrized \\\hline
Self-Dual \cite{Good:2017kjr} & $\infty$ \\\hline
da Vinci-betaK \cite{Good:2018aer,Good:2023ncu} & $\infty$ \\\hline
Schwarzschild-Planck \cite{Good:2019tnf,Good:2020fsw,Good:2023ncu,Moreno-Ruiz:2021qrf,Ievlev:2023xzv} & $\infty$ \\\hline
Fermi-Dirac \cite{Ievlev:2023akh} & $\infty$ \\
\hline
{\bf Worldline Eq.~(\ref{trajectory})} & finite $\&$ parametrized \\
\hline
\end{tabular}
\caption{There are a handful of asymptotically resting worldlines. 
The following list chronologically summarizes the known trajectories possessing asymptotic rest with solved Bogolyubov coefficients. Here, `unparametrized' means the energy production (for Arctx) has not been expressed with a parameter that characterizes the distance traveled. See the parametrized energy, Eq.~(\ref{analytical_energy}), for worldline Eq.~(\ref{trajectory}). } 
\label{dualitytable} 
\end{table} 
\endgroup

\section{Electron in a box}\label{sec:electronbox}
\begin{figure}[h]
    \centering
    \includegraphics[scale=0.25]{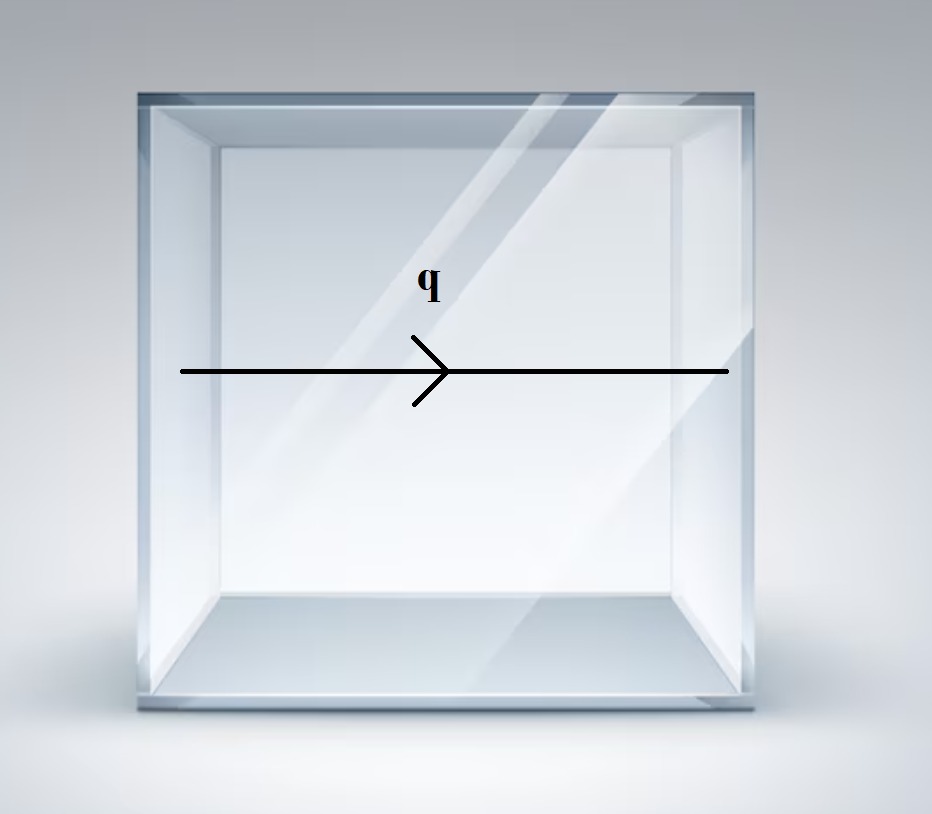}
    \caption{A moving charge, Eq.~(\ref{trajectory}), confined to a geometric cubic volume of space, a `box' with transparent sides (i.e. for our theoretical discussion it is only a mathematical boundary, not an actual physical box, though the figure is suggestive of possible experimental set-up that would involve actual boundaries).  The minimal length of the box is $L = 4 r/\kappa$, where $r$ is a dimensionless free parameter, and $\kappa$ is the dimensionful scale of the system with units of acceleration. The charge radiates thermal photons distributed according to a one-dimensional Planck spectrum, Eq.~(\ref{planck}), shown explicitly when $r=1$ in the non-relativistic regime.  For large box sizes, $r\to\infty$ (all speeds), thermal radiation is implicit via energy emission, Eq.~(\ref{rapidityElargeR}).  } 
    \label{box_fig}
\end{figure}

We start with an ansatz for the equation of motion of the moving point charge.  Assume the charge, such as an electron, travels along a particular worldline confined to a finite space. Allow it to accelerate rectilinearly within this finite distance (confined to a `box'; see Fig. \ref{box_fig}), where the worldline is:
\begin{equation}
    t(x)=x\left(\frac{1}{s}-\frac{1}{r}\right)+\frac{2}{\kappa}\tanh^{-1}\left(\frac{\kappa x}{2 r}\right).
    \label{trajectory}
\end{equation}
The electron, moving in a straight line, traverses the distance from $x=-2r/\kappa$ to $x=+2r/\kappa$, where $x$ is the single space coordinate, $0<s<1$ is the maximum speed of the electron, $t$ is the dependent variable and measure of lab time, $\kappa$ is the dimensionful scale of the system with units of acceleration, and $r>0$ is a dimensionless free parameter introduced to characterize displacement conveniently. 

\begin{figure}[h]
    \centering
    \includegraphics[scale=0.9]{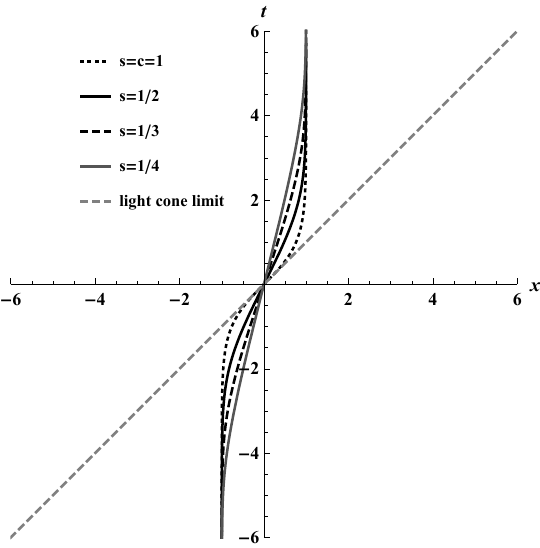}
    \caption{ Trajectories, $t(x)$, from Eq. (\ref{trajectory}) at $r=1$ with different values of maximum speeds $s\rightarrow[1/4,1/3,1/2,1]$. The dashed gray line in the plot at $45^\circ$specifies the light cone limit, and, by definition, no time-like trajectory can cross it. An electron moves along the trajectory from $x\to(-1,1)$ at $\kappa=2$ substituted into Eq. (\ref{trajectory}). Notice the finite distance travelled.} 
    \label{trajectory_fig}
\end{figure}

Eq.~(\ref{trajectory}) describes a continuous, globally defined, timelike worldline starting from rest, reaching speed $s$, and returning to rest.  The key trait is finite distance travel. We have plotted Eq.~(\ref{trajectory}) in Fig.~\ref{trajectory_fig} with $\kappa=2$ and $r=1$, depicting $x\to (-1,1)$.
The figure shows the trajectories for different maximum speeds, $s$. These paths travel a finite distance for all values of $0<s<1$.  In the causal limit $s\to 1$, the charge takes the least time to cover the distance. 


\subsection{Dynamics, \textit{t}(\textit{x})}
\begin{figure}[h]
    \centering    \includegraphics{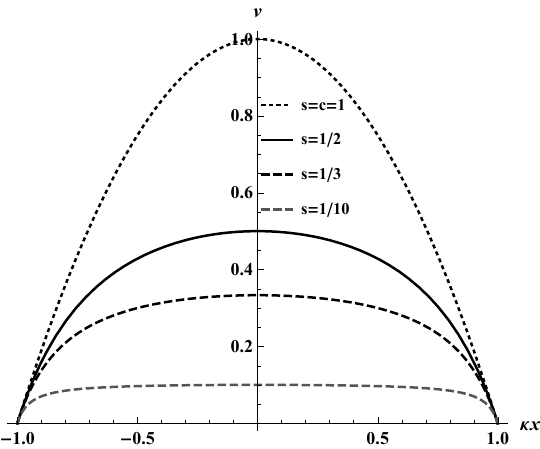}
    \caption{A plot of velocity versus distance (scaled by $\kappa$) traveled, $v$ vs $\kappa x$, for different peak velocities, $s$, at $r=0.5$ which ensures the traversal of $\kappa x$ from $(-1,1)$ for visual simplicity, since $\kappa x\to (-2r,2r)$ depicting the asymptotic resting situation from Eq. (\ref{speed}). The maximum speed is reached at the spatial origin, $v(0) = s$. Notice the smallest speed, $s=0.1$, differs from the smallest speed in Fig. \ref{trajectory_fig}. The smaller the speed, the flatter the velocity curve, consistent with an intuition of thermal equilibrium based on time dependence shown by the dashed gray line at $s=1/10$.} 
    \label{speed_fig}
\end{figure}

\begin{figure}
    \centering
    \includegraphics[width=\columnwidth]{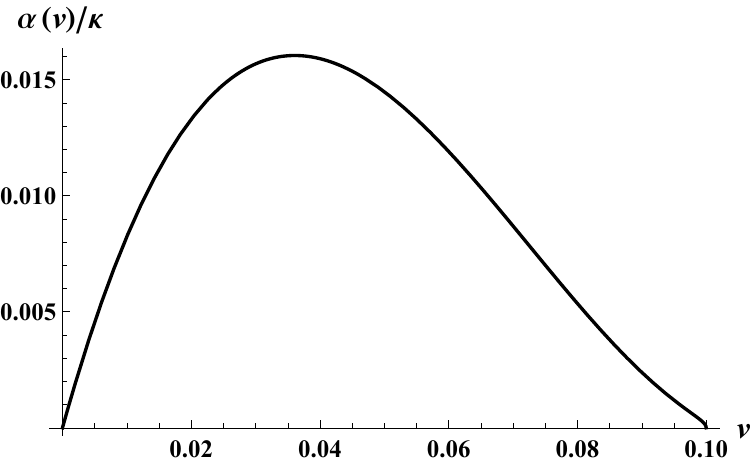}
    \caption{ Plot of the acceleration, $\alpha(v)$ (scaled by $\kappa$) vs velocity $v$ from Eq.~(\ref{alpha}) for max speed $s=0.1$ (non-relativistic regime) and box size $r=1$. The $\alpha(0)=0$ value depicts the electron with zero acceleration at the starting position at the left edge of the box. The acceleration of the electron with velocity $v$ increases to its max speed $v=0\to s$, which occurs at the spatial origin of the box. Subsequently and symmetrically, the acceleration turns negative (not pictured), and the electron's velocity slows down, decreasing to zero as $v=s \to 0$ from the spatial origin to the right side of the box. See also Fig.~\ref{trajectory_fig}.}
    \label{alpha_fig}
\end{figure}

The proper acceleration of the electron is useful for understanding the motion and is necessary to calculate the power and total energy emitted. The proper acceleration is, see, e.g., \cite{Good:2017ddq}
\begin{equation}
    \alpha(x)=\frac{\diff{}}{\diff{x}}\gamma(x).
    \label{general_alpha}
\end{equation}
Here the Lorentz factor is $\gamma(x)$=$1/\sqrt{1-({\diff{t}/\diff{x}})^{-2}}$. We obtain $\diff{t}/\diff{x}$, by taking the derivative of Eq.~(\ref{trajectory}):
\begin{equation}
    \frac{\diff{t}}{\diff{x}}=\frac{1}{s}+\frac{1}{r}\frac{1}{(2r/\kappa x)^2 - 1}.
    \label{dt/dx}
\end{equation}
Inverting Eq.~(\ref{dt/dx}), gives the velocity, $v(x)$, for the trajectory,  
\begin{equation}
    v(x)=s\frac{4r^{2}-(\kappa x)^2}{4r^2-(\kappa x)^2(1-s/r)}.
    \label{speed}
\end{equation}
Notice when $r \to \infty$ then $v(x) \to s$.  Notice also that when $s\ll 1$, then $v(x) \approx s$. On plotting Eq.~(\ref{speed}) in Fig.~\ref{speed_fig}, we get the expected behavior, which starts from rest $v(-2r/\kappa)=0$, reaches a maximum speed, $v(0) = s$, and comes back to rest $v(+2r/\kappa)=0$ after covering a finite distance.  
Expressed as a function of velocity $v$, the proper acceleration $\alpha(v)$ is:
\begin{equation}
    \alpha(v)=
   \pm \kappa  v \gamma^3 \left(1-\frac{v}{s}\right)^{1/2} \left(1 - \frac{v}{s}+\frac{v}{r}\right)^{3/2},
    \label{alpha}
\end{equation}
It is also straightforward to use Eq.~(\ref{speed}) to find the Lorentz factor and obtain the 
proper acceleration $\alpha(x)$ from Eq.~(\ref{general_alpha}). 
On plotting $\alpha(x)$ of Eq.~(\ref{general_alpha}) in Fig. \ref{alpha_fig}, we see that our electron reaches a maximum speed, $s$, when $\alpha=0$ on the graph. For max speeds, $s \to 1$, $\alpha \to \pm \infty$.   
\subsection{Larmor Power, \textit{P}(\textit{v})}
\begin{figure}[h]
    \centering
    \includegraphics{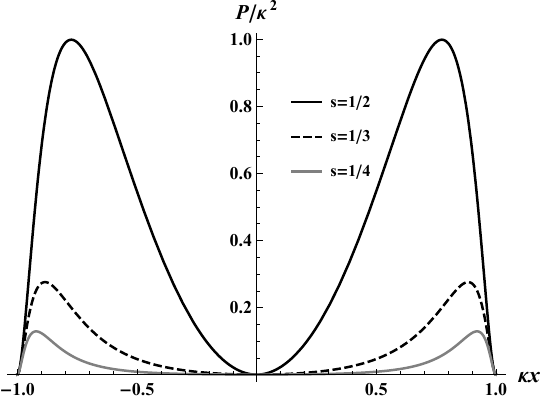}
    \caption{Plots of Larmor power, Eq. (\ref{power}), as a function of distance, $P/\kappa^2$ vs $\kappa x$, for different values of maximum electron speeds, $s\rightarrow$[$1/2,1/3,1/4$]. Here, $r=0.5$ to give the distance from the origin as one. For visual simplicity, we have normalized the vertical axis with the maximum power, $P_{\textrm{max}}/\kappa^2$, for $s=1/2$.  
    Most of the power is emitted near the edges of the box.}
    \label{power_fig}
\end{figure}

 The power emitted by the electron is given by the Larmor-Liénard relation, parametrized by velocity, $v$, as the independent variable:
\begin{equation}
    P(v)=\frac{e^2}{6\pi}\alpha(v)^2.
    \label{general_power}
\end{equation}
With $\alpha(v)^2$ from Eq.~(\ref{alpha}), the power is 
explicitly expressed in terms of velocity, $v$, as:
\be P(v) = \frac{e^2}{6\pi}\kappa^2  v^2 \gamma^6 \left(1-\frac{v}{s}\right) \left(1 - \frac{v}{s}+\frac{v}{r}\right)^{3}.\label{power}\ee
We have plotted the results of the Larmor power as a function of distance $P(x)$ in Fig.~\ref{power_fig}, which demonstrates a two-lobe power trend minimizing at $x=0$ where the electron's speed reaches $s$, then becomes $P = 0$ when the electron comes to rest. 

\subsection{Energy from Power, \textit{E}(\textit{r},\textit{s})}
\begin{figure}[h]
    \centering
    \includegraphics[width=0.9\columnwidth]{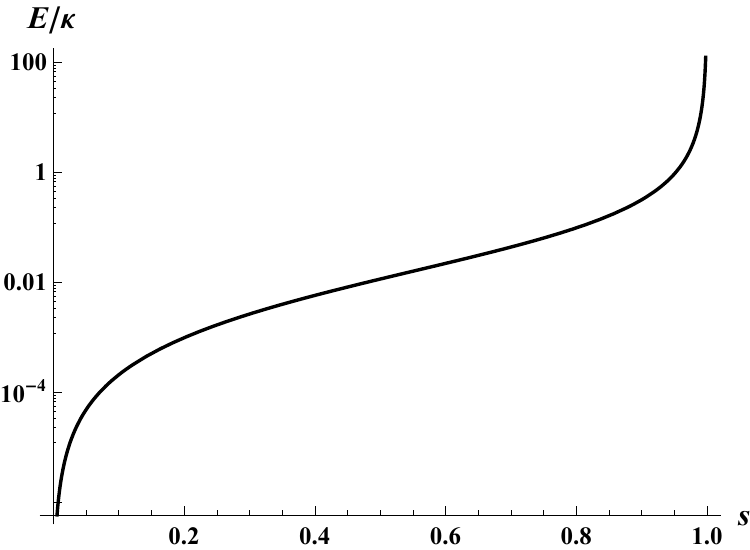}
    \caption{The plot of total energy, $E/\kappa$ as a function of maximum speed, $s$ for $r=1$. The plot is produced from Eq. (\ref{energy_integral}) from Larmor power $P(v)$; or equivalently using the energy of Eq. (\ref{I(w)_energy}),  from $I(\omega)$. The key takeaway is larger maximum speeds result in exponentially more energy production.}
    \label{E_s_fig}
\end{figure}

We obtain the total finite energy emitted by the electron via integrating the Larmor power, $P$, from Eq.~(\ref{general_power}) over time. Using the appropriate Jacobian to change the integration variable from $t\rightarrow x$, we confine the electron to its integration bounds:
\begin{equation}
    E=\frac{e^2}{6\pi}\int_{-2r/\kappa}^{+2r/\kappa}\alpha(x)^{2}\frac{dt}{dx}dx.
    \label{energy_integral}
\end{equation}
The limits of the integral are due to the finite displacement of the trajectory of our electron coming to rest at the edges of the box. We substitute $\diff{t}/\diff{x}$ by Eq.~(\ref{dt/dx}), and analytically integrate over the space from $x=-2r/\kappa$ to $x=+2r/\kappa$. The total energy is only a function of $r$ and $s$ (size and speed). A plot of the result, Eq.~(\ref{analytical_energy}), is given in Fig. \ref{E_s_fig} with fixed $r=1$, depicting finite emission energy as a function of $s$.  It illustrates the energy divergence as $s\rightarrow 1$.

Despite its lengthiness, we express the energy $E(r,s)$, Eq.~(\ref{energy_integral}), and its dependence on the box size, $r$, and maximum electron speed, $s$, for generality.
Using $\gamma_s = 1/\sqrt{1-s^2}$, the answer is:
\be E(r,s) = \frac{e^2\kappa}{12\pi}\left(\frac{ s \gamma_s^2}{4 r} - 1  + E_+(r,s) + E_-(r,s)\right),
\label{analytical_energy}
\ee
where $E_\pm$ are defined by,
\be \nonumber E_\pm = A_\pm \tanh^{-1}\Phi^{-1/2},\quad \Phi = \frac{r(s\mp 1)}{r(s\mp 1)\pm s},\ee
and $A_\pm$, $B_\pm$, $C_\pm$ are as follows:
\be\nonumber  A_\pm = B_\pm\left(2+\frac{3 s^2}{8r^2} - 2\frac{s}{r} +C_\pm\right),\ee
\be\nonumber  B_\pm = \frac{\Phi^{1/2}}{(s\mp 1)^2}, \ee
\be \nonumber C_\pm = \pm \frac{5}{4 r} \mp \frac{1}{s} \mp s \mp \frac{s}{4r^2} \pm \frac{3}{4}\frac{s^2}{r}. \label{TOTAL_ENERGY}\ee
When the electron is slow, Eq.~(\ref{analytical_energy}), to leading order is
\be E(s) = \frac{e^2\kappa s^2}{36\pi}.\label{NR_energy}\ee
We will prove this is the radiated amount of thermal energy released (only valid in the non-relativistic regime).  But first, let us look at the limit of large displacements. 

\begin{figure}
    \centering
    \includegraphics[width=\columnwidth]{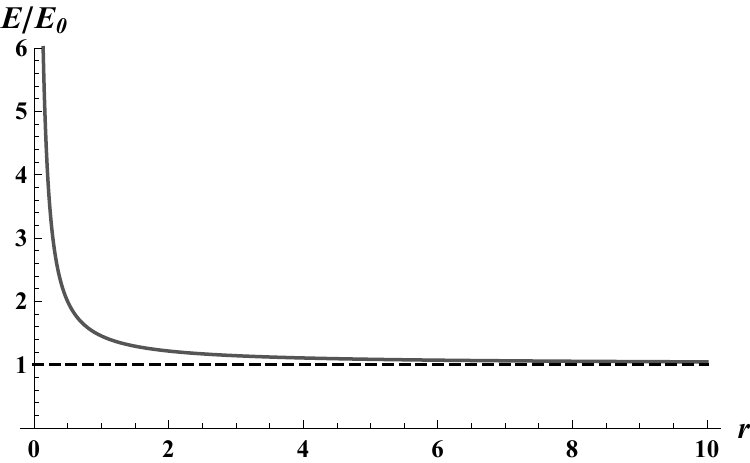}
    \caption{Total energy as a function of the distance traveled, $r$, by plotting Eq. (\ref{analytical_energy}) with $\kappa$; ${E}/{\kappa}$. The black dashed line shows the asymptotic behavior of the energy followed to large $r$. For visual clarity, the energy has been normalized by the minimum energy $E_0 = (\eta/s -1)\kappa e^2/12\pi$, from Eq. (\ref{rapidityElargeR}), which is the energy in the large $r$ limit for $s=1/4$.  The key takeaway is that the energy does not asymptotically drop to zero but instead approaches a finite value, Eq. (\ref{rapidityElargeR}). This is a signature of thermality.}
    \label{energy_r_fig}
\end{figure}

For large distances traveled, $r\rightarrow \infty$ limit, the energy $E(r,s)$ of Eq.~(\ref{analytical_energy}) collapses to the following expression, in terms of final rapidity $\eta = \tanh^{-1} s$:
\begin{equation}
    \lim_{r\rightarrow \infty}E(r,s)=\frac{e^2\kappa}{12\pi}\left(\frac{\eta}{s}-1\right).
    \label{rapidityElargeR}
\end{equation}
The result shows a finite energy emission for the electron in the large distance limit; see Fig \ref{energy_r_fig}.  This is a critical result of this paper.  This energy, Eq.~(\ref{rapidityElargeR}), has the same dependence on final speed as the electron created during beta decay \cite{PhysRev.76.365,Zangwill:1507229,Good:2022eub, Jackson:490457}. 

Because beta decay has recently been shown to be thermal \cite{Ievlev:2023inj} and measured \cite{ptep,RDKII:2016lpd} as such, Eq.~(\ref{rapidityElargeR}) suggests the radiation from Eq.~(\ref{trajectory}) may be thermal in the large displacement limit, i.e., large boxes may contain CATs. However, to convincingly demonstrate a CAT, we must show that the spectrum has a Planck factor. We compute the spectrum for a particular box size in the following section, Section \ref{sec:spectrum}, and specialize to small speeds in Section \ref{sec:CAT} to look for the Planck factor.

\section{Spectral Analysis}\label{sec:spectrum}
\subsection{Spectral Distribution, d\textit{I}(\textit{ω})/dΩ  }
The spectral distribution for the straight-line traveling electron is calculated using the formula, see e.g., Eq.~(14.70) of \cite{Jackson:490457} or Eq.~(23.89) of Zangwill \cite{Zangwill:1507229}:
\begin{equation}
    \frac{\diff{I(\omega)}}{\diff{\Omega}}=\frac{\omega^{2}}{16\pi^{3}}\sin^{2}\theta\left|j(\omega,k_{x})\right|^{2},
    \label{I_general}
\end{equation}
where $j(\omega,k_{x})$ is the current and $k_x=\omega\,\cos(\theta)$. The quantity $j(\omega,k_{x})$ can be computed by:
\begin{equation}
    j(\omega,k_{x})=e\int_{-2r/\kappa}^{2r/\kappa}e^{i\phi} dx,
    \label{j_current_general}
\end{equation}
where $\phi=k_{x}x-\omega t(x)$. Then using Eq. (\ref{trajectory}), we get: 
\begin{equation}
    \phi/\omega=x \cos\theta-x\left(\frac{1}{s}-\frac{1}{r}\right) -\frac{2}{\kappa}\tanh^{-1}\left(\frac{\kappa x}{2r}\right),
    \label{phi}
\end{equation}
where we have substituted in expression $k_x=\omega\cos\theta$.  The integral in Eq.~(\ref{j_current_general}) with $r=1$ can be performed with a substitution. Finally, complex conjugating the current gives: 
\begin{equation}
   \left|j(\omega,k_x)\right|^{2}= \frac{16\pi^2 e^2 \omega^2}{\kappa^{4}} \csch^{2}\left(\frac{\pi \omega}{\kappa}\right)\left|M\right|^{2},
   \label{j_square}
\end{equation} 
where $M$ is the hyper-geometric Kummar function, $_1F_1$, dependent on $s$, the max speed of the electron, $\theta$ the polar angle, and $\omega$ the photon frequency. Explicitly,
\be 
M = {}_1F_1\left(
    1-\frac{i \omega}{\kappa}, 2, \frac{4 i \omega}{\kappa}(1- \frac{1}{s} + \cos\theta)
    \right).\label{kummer}\ee
The spectral distribution of Eq.~(\ref{I_general}) is, therefore,

\be 
 \frac{\diff{I(\omega)}}{\diff{\Omega}}=\frac{e^2 \omega^{4}}{\pi\kappa^{4}} \sin^{2} \theta \csch^2 \left(\frac{\pi \omega}{\kappa}\right)\left|M\right|^{2}.
 \label{spec_dis_ris1}
\ee
The spectral distribution, Eq.~(\ref{spec_dis_ris1}), which has set $r=1$, is a good starting point for investigating thermality, which we do in Section \ref{sec:CAT} for low speeds. However, in the following subsection, we will first integrate over the solid angle to get the $I(\omega)$ spectrum relevant for all speeds $0<s<1$. 
\subsection{Spectrum, \textit{I}(\textit{ω})}

\begin{figure}[h]
    \centering
    \includegraphics[width=\columnwidth]{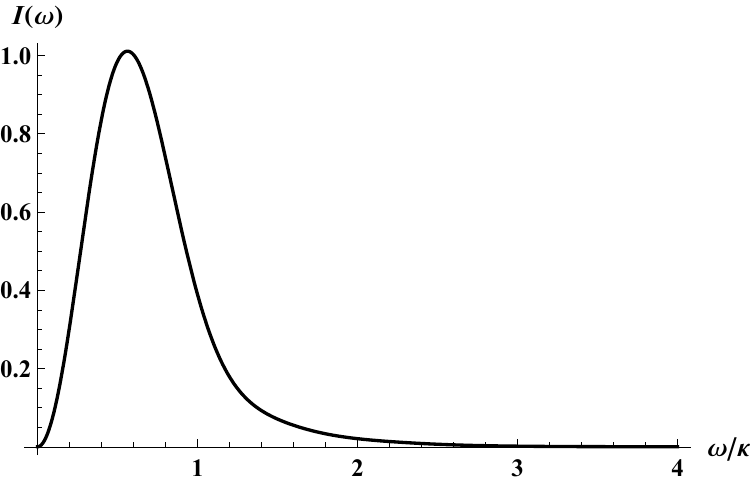}
    \caption{A plot of the spectrum, $I(\omega)$ vs $\omega/\kappa$ trend obtained 
 from Eq.~(\ref{spectrum_integral}) for $s=0.4$ asymptotically approaching $0$ as $\omega\to \infty$. The spectrum $I(\omega)$ is normalized by its maximum value; $I(\omega)$=$I/I_{\textrm{max}}$, where here $I_{\textrm{max}}$ is the maximum of $I(\omega)$ when $s=0.4$. Notice there is no infrared divergence, and the shape is qualitatively similar to a $3+1$ dimensional Planck curve. } 
    \label{spectrum_r=1}
\end{figure}

We obtain the spectrum $I(\omega)$ by integrating over the solid angle, $\diff{\Omega}=\sin\theta \diff{\theta} \diff{\phi}$.  The general prescription uses the current by integrating Eq. (\ref{I_general}) employing Eq. (\ref{j_square}):
\begin{equation}
  I(\omega)=\frac{2 e^2 \omega^{4}}{\kappa^{4}}\csch^{2}\left(\frac{\pi \omega}{\kappa}\right)\int_{0}^{\pi}\sin^{3}\theta\left|M\right|^{2}d\theta.\label{spectrum_integral}    
\end{equation}
In general cases, the $I(\omega)$ spectrum is not normally analytic, and Eq.~(\ref{spectrum_integral}) is no exception. However, as we will show, an analytic form for the spectrum exists at slow speeds, Eq.~(\ref{NR_spectrum}). We have plotted the spectrum $I(\omega)$, Eq. (\ref{spectrum_integral}) in Fig. \ref{spectrum_r=1}.  Notice the lack of an infrared divergence, which is expected for an asymptotically static trajectory, unlike asymptotic constant velocity trajectories like the beta decay trajectory, see, e.g., \cite{Lin:2024ihr,ptep}.

\subsection{Energy from Spectrum, \textit{E}(\textit{s})}
To compute the total energy of the classical radiation emitted by the electron, $E_{\textrm{total}}$, one integrates Eq.~(\ref{I_general}) over the solid angle, $\diff{\Omega}$ and frequency, $\diff{\omega}$ via:
\begin{equation}
    E_{\textrm{total}}=\int_0^\infty \frac{\diff{I(\omega)}}{\diff{\Omega}} \diff{\Omega}\diff{\omega} =\int_{0}^{\infty}I(\omega)  \diff{\omega},\label{E_total_I_general}
    \end{equation}
Using the spectra $I(\omega)$ from Eq.~(\ref{spectrum_integral}), we obtain energy emitted, remembering that $r=1$, as:
\begin{equation}
    E_{\textrm{total}}^{r=1}(s)=\int_{0}^{\infty}\int_{0}^{\pi}\frac{2 e^2 \omega^{4}}{\kappa^{4}}\csch^2(\frac{\pi \omega}{\kappa}) \sin^{3}\theta \left|M\right|^{2} d\theta d\omega. 
    \label{I(w)_energy}
\end{equation}
Numerical integration of the above equation gives $E_\textrm{total}$ as a function of max speed, $s$. The total energy of Eq.~(\ref{I(w)_energy}) checks with the energy Eq.~(\ref{analytical_energy}) obtained from the Larmor power, verifying consistency and the correctness of the $r=1$ spectral distribution Eq.~(\ref{spec_dis_ris1}). 
\section{Classical Acceleration Temperature}\label{sec:CAT}
This section will analyze the spectral distribution, Eq.~(\ref{spec_dis_ris1}), (which assumes box size $r=1$) and demonstrate thermality in the non-relativistic regime, $s\ll 1$. However, first, we will look at the character of the acceleration in the relevant limits to understand and motivate the situation from a temporal-spatial perspective.  

\subsection{Peel Acceleration, $|\mathcal{P}| = \kappa$}
We show that the peel is constant to first order in large box sizes or small speeds. This result hints at thermality and suggests Eq.~(\ref{spec_dis_ris1}), valid for $r=1$, and possessing no apparent Planck factor, may be thermal in the regime where speeds are low, i.e. $s\ll 1$. 

Consider the Carlitz-Willey trajectory \cite{carlitz1987reflections}:
\be V = \frac{1}{\kappa} e^{\kappa U},\qquad U = t - x.\label{CW_trajectory}\ee
This expression is written in light-cone coordinates where $V= t+x$. The thermal object of interest associated with acceleration is the peel, e.g., \cite{carlitz1987reflections,Bianchi:2014qua,Barcelo:2010pj,Ievlev:2023xzv}. One can use light-cone coordinates, $(V,U)$, and express the peel of the above Carlitz-Willey trajectory.  The constant peel gives the temperature of the quantum radiation from the moving mirror:
\be \kappa = \frac{V''(U)}{V'(U)},\qquad  T_{\textrm{mirror}} = \frac{1}{2\pi} \frac{V''(U)}{V'(U)} = \frac{\kappa}{2\pi}.\ee 
Create a similar peel by considering the following equation of motion for a moving electron in the context of beta decay \cite{ptep}, 
\be x = \frac{s}{\kappa} e^{\kappa u_s},\qquad u_s = t - \frac{x}{s},
\label{diff_trajectory}
\ee
where $s$ is the maximum speed of the electron.  This trajectory resembles the eternal thermal Carlitz-Willey moving mirror in the sense that the peel in the coordinates $(x,u_s)$ is constant, which gives the temperature:
\be \kappa = \frac{x''(u_s)}{x'(u_s)}, \qquad  T_{\textrm{electron}} = \frac{1 }{2\pi} \frac{x''(u_s)}{x'(u_s)} = \frac{\kappa}{2\pi}.\label{peel}\ee 
Notice that $u_s$ is not the usual light-cone coordinate but a `stretched-space' retarded time coordinate.

Now that the expression for peel is appropriately motivated let us compute the first expression of Eqs.~(\ref{peel}) for our boxed trajectory, Eq.~(\ref{trajectory}),
\be \mathcal{P} \equiv \frac{x''(u_s)}{x'(u_s)} =\pm \kappa \left[1+\frac{1}{r}\left(\frac{1}{v}-\frac{1}{s}\right)^{-1}\right]^{3/2},\label{peel_box}\ee
where $v$ is the velocity, Eq.~(\ref{speed}), of the trajectory Eq.~(\ref{trajectory}).  It is readily seen that for large box sizes $r$,
\be \mathcal{P} = \pm \kappa + \mathcal{O}(1/r).\label{peelBOX}\ee
Like the energy result, Eq.~(\ref{rapidityElargeR}), obtained in the limit of large $r$, the dynamic (temporal-spatial) result, Eq.~(\ref{peelBOX}), also suggests thermality is present for large $r$. It should be noted that for small velocities $s \to 0$, Eq.~(\ref{peel_box}) gives
\be \mathcal{P} = \pm \kappa + \mathcal{O}(s)\label{peelBOX1},\ee
which suggests thermality is present for non-relativistic speeds. 

In summary, our trajectory, Eq.~(\ref{trajectory}), in the limit of $r \to \infty$ or $s\ll 1$, may exhibit thermality as given by the temperature, $T=\kappa/2\pi$, which is a conjecture by analogy with the temperatures of the moving mirror trajectory Eq.~(\ref{CW_trajectory}), and the electron trajectory of beta decay, Eq.~(\ref{diff_trajectory}).
\subsection{Proper Acceleration, \textit{α}(\textit{v})}
The proper acceleration, Eq.~(\ref{alpha}), of trajectory Eq.~(\ref{trajectory}) is identical to the thermal acceleration exhibited during beta decay, $\alpha_0(v) = \kappa v \gamma^3(1-v/s)^2$ \cite{Good:2022eub}, in the leading order for a large box.  This is consistent, as it should be, with the total energy emitted shown in  Eq.~(\ref{rapidityElargeR}), which assumed a large finite displacement. 

It is straightforward to see the leading order of Eq.~(\ref{alpha}) in terms of large $r$,
\be \alpha(v) =  \alpha_0(v) + \mathcal{O}(1/r).\ee
In addition to expanding in terms of large $r$, one may also expand to leading order in small speeds $s$,
\be \alpha(v) =  \frac{\kappa v^3 \gamma^3}{s^2} + \mathcal{O}(1/s).\label{low_speeds_proper_acc}\ee
This result is identical to low speeds, $s\ll 1$, of beta decay acceleration, $\alpha_0(v) = \kappa v \gamma^3(1-v/s)^2 \approx \kappa v^3 \gamma^3 / s^{2}$, which suggests an electron moving along Eq.~(\ref{trajectory}) in the non-relativistic regime, emits thermal radiation.  Taken together, Eq.~(\ref{peelBOX1}) and Eq.~(\ref{low_speeds_proper_acc}), properly motivate a study of the spectrum.  At low speeds, one expects a Planck distribution, as given by Eq.~(\ref{NR_spectrum}).  
\subsection{Planck's Law, \textit{r} = 1 \& \textit{s} $\ll$ 1}
The spectral distribution, Eq.~(\ref{spec_dis_ris1}), assumes box size $r=1$ and may demonstrate thermality in the non-relativistic regime, $s\ll 1$. At these low speeds, we can take the spectral distribution to leading order in $s^2$ and integrate over the solid angle to get the spectrum: 
\be I(\omega) = \frac{e^2 s^2 }{3 \pi^2} \frac{2\pi\omega/\kappa}{e^{2 \pi  \omega/\kappa} -1}  + \textrm{O.T.'s}.\label{NR_spectrum}\ee
The oscillation terms, $\textrm{O.T.'s} = X Y$, the product of a complicated function
\be X(\omega) = \frac{\csch^2\pi  \omega /\kappa}{32e^{4\pi \omega/\kappa}}\left[\frac{\sin{(4\omega/\kappa)}}{4\omega/\kappa}-\cos{(4\omega/\kappa)}\right]e^{2\pi \omega/\kappa}.\label{OT1}\ee
and another, also complicated, $s$-dependent function,
\be Y(\omega,s) = -\frac{2^{\frac{4 i \omega }{\kappa }}  e^{-\frac{4 i \omega }{\kappa }}}{\Gamma \left(\frac{i \omega }{\kappa }+1\right)^2}s^2 e^{\frac{4 i \omega }{\kappa s }} \left(-\frac{i\omega }{ \kappa  s}\right)^{\frac{2 i \omega }{\kappa }}  + \textrm{h.c.} \label{OT2}\ee
An integral of $I(\omega)$, Eq.~(\ref{NR_spectrum}), however, gives a surprisingly clean result:  
\be E = \int_0^\infty I(\omega) \diff{\omega} = \frac{e^2\kappa s^2}{36\pi},\ee
which agrees with Eq.~(\ref{NR_energy}), confirming the form of the spectrum, Eq.~(\ref{NR_spectrum}). 

Interestingly, neglecting the oscillation terms responsible for IR-finite regularization is possible in the above integral, as they do not spoil the Planck distribution; see a similar situation in \cite{Ievlev:2023xzv}.  That is, we can use first term in the spectrum Eq.~(\ref{NR_spectrum}), named $I_0(\omega)$, to show that: 
\be I_0(\omega) = \frac{e^2s^2 }{3 \pi^2} \frac{2\pi\omega/\kappa}{e^{2 \pi  \omega/\kappa} -1}, \quad E = \int_0^\infty I_0(\omega) \diff{\omega} = \frac{e^2\kappa s^2}{36\pi},\label{spectrum_low_speeds}\ee
demonstrating that the oscillation terms do not contribute to the total radiation energy.
The temperature in $I_0(\omega)$ of Eq.~(\ref{spectrum_low_speeds}), reinstating Kelvin SI units,
\be T = \frac{\hbar \kappa}{2\pi c k_B}, \label{temp}\ee
is independent of max speed $s$ or polar angle $\theta$.  This is in contrast to other cases like the asymptotic static trajectory for complete evaporation associated with the Schwarzschild-Planck system in \cite{Ievlev:2023xzv}, the remnant trajectories in \cite{Lin:2024ihr}, or the eternal black hole trajectory in the appendix of \cite{Ievlev:2023inj}. Notice that $\hbar$ is required for all temperatures measured in Kelvin \cite{SI}. Its appearance in this classical computation does not signify any use of quantum theory, i.e., Eq.~(\ref{temp}) is a CAT.  

\section{Particle Production}\label{sec:particles}

\subsection{Particle Spectrum, \textit{N}(\textit{ω})}
We can compute the particle spectrum $N(\omega)$, noticing the quantum inclusion of $\hbar$,
by using the following relation:
\be N(\omega) = \frac{I(\omega)}{\hbar \omega}.\label{particle_cout_formula} 
\ee
That is, the particle spectrum can be calculated as an integral of the spectral distribution:
\be
    N(\omega)=\int_{0}^{\pi}\int_{0}^{2\pi}\frac{1}{\hbar \omega}\frac{\diff{I}}{\diff{\Omega}}\sin\theta  \diff{\theta} \diff{\phi}.
    \label{number_particles_general}
\ee
This assumes each particle carries a quantum of energy $\hbar \omega$, where the idea of photons in classical electrodynamics is considered semi-classical \cite{Jackson:490457}.

\subsection{Fixed Box Size, \textit{r} = 1}
\begin{figure}
    \centering    \includegraphics[width=\columnwidth]{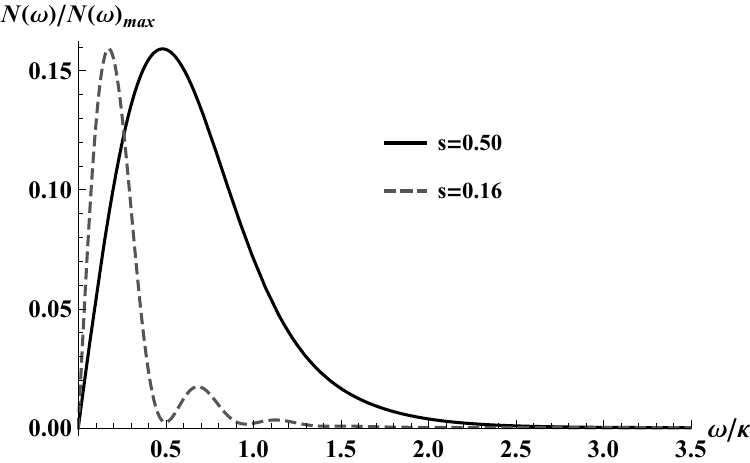}
\caption{$N(\omega)/N(\omega)_{\textrm{max}}$ vs $\omega/\kappa$ maximum speeds, $s\to (0.50,0.16)$ via the numerical solution of Eq. (\ref{number_particles_r=1}), at $r=1$. Here $N(\omega)$ is normalized to its peak value: $N(0.47)=1.08\times 10^{-2}$ at $s = 0.5$ and $N(0.176)=7.26\times 10^{-2}$ at $s = 0.16$ found from the plotted data points. Notice the appearance of a tail oscillation due to lower maximum speed (dashed line). These oscillations are critical for regularizing finite particle count but unnecessary for computing the radiated energy. }
    \label{fig:particle_spectrum}
\end{figure}

We specialize to box size $r=1$ and use Eq.~(\ref{spec_dis_ris1}) in Eq.~(\ref{number_particles_general}), integrate over $\phi$, and obtain the particle spectrum $N(\omega)$ as a function of frequency $\omega$, 
\begin{equation}
    \label{number_particles_r=1}
    N(\omega)=\int_{0}^{\pi} \frac{2 e^2 \omega^{3}}{\hbar \kappa^{4}}\csch^2(\frac{\pi \omega}{\kappa})\sin^{3}\theta |M|^{2} \diff{\theta}.
\end{equation} 
This integral cannot be done analytically.  Nevertheless, numerically evaluating Eq.~(\ref{number_particles_r=1}), we obtain the particle spectrum in Fig. \ref{fig:particle_spectrum}.  The fact that the worldline comes to a stop is the dynamic reason the particle spectrum $N(\omega)$ suffers no IR-divergence, see e.g. \cite{Good:2022eub,ptep,Ievlev:2023inj}. To better understand Eq.~(\ref{number_particles_r=1}), we need to specialize to non-relativistic speeds, which we do in the next section.  

\subsection{Slow Max Speed, \textit{r} = 1 \&   \textit{s} $\ll$ 1}
Low speeds are characterized by $s\ll 1$ where we still have $r=1$.  Eq.~(\ref{particle_cout_formula}) is succinctly reformulated using the $I_0(\omega)$ in Eq.~(\ref{spectrum_low_speeds}) as follows:  
\be  N_0(\omega)= \frac{2}{3\pi}\frac{e^2 s^2}{\hbar \kappa} \frac{1}{e^{2\pi \omega/\kappa}-1}.\label{planck}\ee
We have neglected the oscillation terms because we have used Eq.~(\ref{spectrum_low_speeds}) and not the full expression Eq.~(\ref{NR_spectrum}) here, but as can be seen, they are precisely the needed ingredients to regularize the infrared divergence. In contrast to energy, the finite total particle count depends on the oscillations. This is because  
\be N_0(\omega) = \frac{I_0(\omega)}{\hbar \omega},\quad N_\infty = \int_0^\infty N_0(\omega) \diff{\omega},\ee
is not finite (IR-divergent), and $N_\infty$ does not count the particles. Instead, the total particle count  
\be N = \int_0^\infty N(\omega) \diff{\omega} = \textrm{finite},\label{N_particlecount}\ee
is convergent and includes the oscillations, via $N(\omega)$ of Eq.~(\ref{particle_cout_formula}). The fact that the energy $E$, Eq.~(\ref{spectrum_low_speeds}), has no dependence on the oscillation terms and the particle numbers do, Eq.~(\ref{N_particlecount}), can mean two different things physically: (a) the oscillation part produces soft particles\footnote{These O.T.'s might rightly characterize infinite `anti-soft' particles that regularize the infinite soft particles of the 1D Planck term.} that do not carry energy, or (b) some particles produced in the oscillations produce positive energy while some negative energy, and they exactly cancel. In \cite{Good:2015nja}, particle numbers do not reveal negative energy flux; thus, it is an open question on how to interpret the particle production associated with the oscillations physically. 

\begin{figure}[h]
    \centering
    \includegraphics[width=0.95\columnwidth]{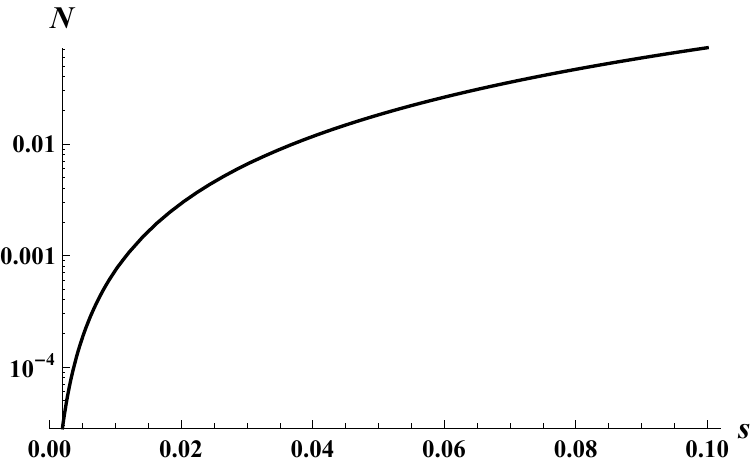}
    \caption{Particle number, $N$, as a function of $s$ in the non-relativistic regime $s\to (0,0.1)$, shows monotonic increase in approach to the relativistic regime for $r=1$.}
    \label{N(s)_fig}
\end{figure}

The particle count results of Eq.~(\ref{N_particlecount}) using Eq.~(\ref{particle_cout_formula}) via Eq.~(\ref{NR_spectrum}) with the oscillation terms, Eq.~(\ref{OT1}) \& Eq.~(\ref{OT2}), in the non-relativistic regime are plotted in Fig. \ref{N(s)_fig}. Notice the faster the maximum speed of the electron, the more particle creation. 

Having derived the photons, let us consider the scalars from the dual mirror moving along Eq.~(\ref{trajectory}). The quantum radiation emitted by an accelerating mirror to the right is found using the mirror-electron mapping recipe see, e.g., \cite{Ievlev:2023inj,Good:2023hsv}.  One finds from Eq.~(\ref{spec_dis_ris1}) with $s\ll 1$ and $r=1$, the scalar particle production beta Bogolybov coefficients,
\be |\beta_{pq}|^2 =\frac{4 s^2}{\kappa\pi} \frac{p q }{(p+q)^3}  \frac{1}{e^{2\pi(p+q)/\kappa}-1} ,\label{betasquared}\ee 
where we have ignored the oscillation terms. The scalars (not photons\footnote{We emphasize that the moving mirror in (1+1)-dimensions is \emph{not} an actual electron in (3+1)-dimensions, though, there is an exact, functional, identity mapping between the two. The point is that the scalar particle spectrum emitted by the (1+1) moving mirror has a one-to-one photon particle spectrum counterpart emitted by the (3+1) ordinary electron. In other words, there is a ``duality'' between the two systems \cite{Nikishov:1995qs}, a classical-quantum correspondence whose physical content in 3d space is encoded on a line.}) carry energy emitted to both sides of the mirror, analog-consistent with Eq.~(\ref{NR_energy}), (see, e.g., the general form via Eq. (43) of \cite{Ievlev:2023xzv}),
\be E_{\textrm{mirror}} = \int_0^\infty\int_0^\infty \hbar (p+q) |\beta_{pq}|^2 \diff{p}\diff{q} = \frac{\hbar \kappa s^2}{36\pi}.\ee
In the high-frequency limit \cite{Hawking:1974sw}, $p\gg q$, Eq.~(\ref{betasquared}) gives a result similar to Hawking radiation for particle production, 
\be |\beta_{pq}|^2 = \frac{4 s^2}{\kappa\pi}\frac{q}{p^2}   \frac{1}{e^{2\pi p/\kappa}-1},\label{betasquared_thermal}\ee
where the Planck factor demonstrates $T=\kappa/2\pi$ thermality in frequency $p$, in analog to the late-time Schwarzschild black hole \cite{Hawking:1974rvNATURE}; see also e.g. the late-time Schwarzschild mirror \cite{Good:2016oey} via Eq.~(B3) of \cite{Ievlev:2023inj}, Eq.~3.24 of Carlitz-Willey \cite{carlitz1987reflections}, or Eq.~(1) of Fulling \cite{Fulling_optics}. Eq.~(\ref{betasquared_thermal})'s thermality characterizes the quantum radiation emitted by the moving mirror and establishes the slow-moving electron along Eq.~(\ref{trajectory}) as a thermal analog of black hole evaporation. 

\section{Conclusions}\label{sec:concl}
We have found that a confined, non-periodic, slowly moving, straight-line accelerating electron emits thermal radiation. The key traits are finite particles and energy emission from a finite space. It is experimentally favorable that the thermal effect from the bounded region coincides with the non-relativistic regime. Novel physical results include:
\begin{itemize}
   \item Total energy radiated depends on the finite distance traversed; see Eq.~(\ref{analytical_energy}). 
   \item Total energy radiated does not depend on the finite distance traversed for large travel distances, see Eq.~(\ref{rapidityElargeR}). 
   \item For slow speeds (and fixed box size), the photons are Planck-distributed, Eq.~(\ref{NR_spectrum}). 
\end{itemize}
The total energy, Eq.~(\ref{analytical_energy}), was found analytically in its most general form and is not straightforward. Nevertheless, the simple result is that the amount of energy radiated does not change with distance when the distance is large, Eq.~(\ref{rapidityElargeR}). The result scales as the total energy radiated by the unconfined electron emitting thermal photons during beta decay. Therefore, in the context of Eq.~(\ref{trajectory}), it is natural to conjecture that large boxes have CATs. Still, a Planck distribution cannot be confirmed from the seemingly intractable form of the large-box spectrum.  Regardless, for a fixed travel distance, the slow-moving electron emits Planck-distributed photons; thus, the radiation is demonstrably thermal, Eq.~(\ref{temp}). There is a CAT in the box.
\begin{acknowledgments}
Funding comes partly from the FY2024-SGP-1-STMM Faculty Development Competitive Research Grant (FDCRGP) no.201223FD8824 and SSH20224004 at Nazarbayev University in Qazaqstan. Appreciation is given to the ROC (Taiwan) Ministry of Science and Technology (MOST), Grant no.112-2112-M-002-013, National Center for Theoretical Sciences (NCTS), and Leung Center for Cosmology and Particle Astrophysics (LeCosPA) of National Taiwan University. 
\end{acknowledgments}


\bibliography{main} 
\end{document}